\documentclass[11pt]{article}
\usepackage{utopia}
\usepackage[active]{srcltx}
\usepackage{multicol}
\usepackage{graphicx}
\usepackage{amssymb}
\usepackage{amsthm}
\title{\bf Three-Level Laser Dynamics with the Atoms Pumped by Electron Bombardment}
\author{Fesseha Kassahun\footnote{Email address: fesseha.kassahun@aau.edu.et} \\
                \footnotesize{Department of Physics, Addis Ababa University, P. O. Box 33761, Addis Ababa, Ethiopia}}
                \date{\footnotesize{(Submitted on 25 Sep 2012)}}

\makeatother

\begin{document}
\maketitle
\begin{abstract}
We analyze the quantum properties of the light generated by a three-level laser with a closed cavity and coupled to a vacuum reservoir. The three-level atoms available in the cavity are pumped from the bottom to the top level by means of electron bombardment and we carry out our analysis by putting the noise operators associated with the vacuum reservoir in normal order. The maximum quadrature squeezing of the light generated by the laser, operating far below threshold, is found to be $50\%$ below the coherent-state level. We have also established that the quadrature squeezing of the output light is equal to that of the cavity light and has the same value in any frequency interval. This implies that the quadrature squeezing of the laser light is an intrinsic property of the individual photons.
\end{abstract}
\hspace*{9.5mm}Keywords: Stimulated emission, Photon statistics, Quadrature squeezing, Noiseless\newline
\hspace*{28.5mm}vacuum reservoir
\vspace*{5mm}
\section*{1. Introduction}
 \vspace*{-2mm}
 A three-level laser is a quantum optical system in which light is generated by three-level atoms inside a cavity usually coupled to a vacuum reservoir. In one model of a such laser, three-level atoms initially prepared in a coherent superposition of the top and bottom levels are injected into a cavity and then removed after they have decayed due to spontaneous emission [1,2]. In another model, the top and bottom levels of the three-level atoms injected into a cavity are coupled by coherent light [3,4]. The statistical and squeezing properties of the light generated by three-level lasers have been investigated by several authors [5-11]. It is found that a three-level laser in either model generates squeezed light under certain conditions.
It appears to be quite difficult to prepare the atoms in a coherent superposition of the top and bottom levels before they are injected into the laser cavity. In addition, it should certainly be hard to find out that the atoms have decayed spontaneously before they are removed from the cavity.
On the other hand, the degree of squeezing of the light generated by the three-level laser, with the top and bottom levels coupled by coherent light, is relatively large when the mean photon number is relatively small [4].

Moreover, the quantum analysis of a three-level laser is usually carried out by including the interaction of the atoms inside the cavity with the vacuum reservoir outside the cavity. It may be possible to justify the feasibility of such interaction for a laser with an open cavity into which and from which atoms are injected and removed. However, there cannot be any valid justification for leaving open the laser cavity in which the available atoms are pumped to the top level by electron bombardment. Therefore, the aforementioned interaction is not feasible for a laser in which the atoms available in a closed cavity are pumped to the top level by means of electron bombardment.

We seek here to analyze the quantum properties of the light emitted by the three-level atoms available in a closed cavity and pumped to the top level at a constant rate. Thus taking into account the interaction of the three-level atoms with a resonant cavity light and the damping of the cavity light by a vacuum reservoir, we obtain the photon statistics of the cavity light and the quadrature squeezing of the cavity (output) light. We also determine the quadrature squeezing of the cavity (output) light in any frequency interval. We carry out our calculation by putting the noise operators associated with the vacuum reservoir in normal order and without considering the interaction of the three-level atoms with the vacuum reservoir outside the cavity.

\section*{2. Operator dynamics}

We consider here the case in which $N$ degenerate three-level atoms in cascade configuration are available in a  closed cavity. We denote the top, middle, and bottom levels of these atoms by $|a\rangle_{k}$, $|b\rangle_{k}$, and $|c\rangle_{k}$, respectively. In addition, we assume the cavity mode to be at resonance with the two transitions $|a\rangle_{k}\rightarrow|b\rangle_{k}$ and $|b\rangle_{k}\rightarrow|c\rangle_{k}$, with direct transition between
levels $|a\rangle_{k}$ and $|c\rangle_{k}$ to be dipole forbidden.
The interaction of one of the three-level atoms with the cavity mode can be described at resonance by the Hamiltonian
\begin{equation}\label{1}
\hat{H}=ig\left[(\hat{\sigma}_{a}^{\dagger k}+\hat{\sigma}_{b}^{\dagger k})\hat{b}
-\hat{b}^{\dag}(\hat{\sigma}_{a}^{k}+\hat{\sigma}_{b}^{k})\right],
\end{equation}
where
\begin{equation}\label{2}
\hat{\sigma}_{a}^{k}=|b\rangle_{k}~_{k}\langle a|
\end{equation}
and
\begin{equation}\label{3}
\hat{\sigma}_{b}^{k}=|c\rangle_{k}~_{k}\langle b|
\end{equation}
are lowering atomic operators, $\hat{b}$ is the
annihilation operator for the cavity mode, and $g$ is the coupling constant between the atom and the cavity mode.
We assume that the laser cavity is coupled to a vacuum reservoir via a single-port mirror. In addition, we carry out our calculation by putting the noise operators associated with the vacuum reservoir in normal order. Thus the noise operators will not have any effect on the dynamics of the cavity mode operators. We can therefore drop the noise operator and write the quantum Langevin equation for the operator $\hat{b}$ as
\begin{equation}\label{4}
{d\hat{b}\over dt}=-{\kappa\over 2}\hat{b}-i[\hat{b},\hat{H}],
\end{equation}
where $\kappa$ is the cavity damping constant. Then with the aid of Eq. (\ref{1}), we easily find
\begin{equation}\label{5}
{d\hat{b}\over dt}=-{\kappa\over 2}\hat{b}-g(\hat{\sigma}_{a}^{k}+\hat{\sigma}_{b}^{k}).
\end{equation}

Furthermore, applying the relation
\begin{equation}\label{6}
{d\over dt}\langle\hat{A}\rangle=-i\langle[\hat{A},\hat{H}]\rangle
\end{equation}
along with Eq. (\ref{1}), one can readily establish that
\begin{equation}\label{7}
{d\over dt}\langle\hat{\sigma}_{a}^{k}\rangle=g\langle{\hat{\eta}_{b}^{k}}\hat{b}\rangle
-g\langle\hat{\eta}_{a}^{k}\hat{b}\rangle+g\langle\hat{b}^{\dagger}\hat{\sigma}_{c}^{k}\rangle,
\end{equation}
\begin{equation}\label{8}
{d\over dt}\langle\hat{\sigma}_{b}^{k}\rangle=g\langle{\hat{\eta}_{c}^{k}}\hat{b}\rangle-g\langle\hat{\eta}_{b}^{k}\hat{b}\rangle
-g\langle\hat{b}^{\dagger}\hat{\sigma}_{c}^{k}\rangle,
\end{equation}
\begin{equation}\label{10}
{d\over dt}\langle\hat{\eta}_{a}^{k}\rangle=g\langle\hat{\sigma}_{a}^{\dagger k}\hat{b}\rangle+g\langle\langle\hat{b}^{\dagger}\hat{\sigma}_{a}^{k}\rangle,
\end{equation}
\begin{eqnarray}\label{11}
{d\over dt}\langle\hat{\eta}_{b}^{k}\rangle=g\langle\hat{\sigma}_{b}^{\dagger k}\hat{b}\rangle-g\langle\hat{\sigma}_{a}^{\dagger k}\hat{b}\rangle+g\langle\hat{b}^{\dagger}\hat{\sigma}_{b}^{k}\rangle-g\langle\hat{b}^{\dagger}\hat{\sigma}_{a}^{k}\rangle,
\end{eqnarray}
\begin{equation}\label{12}
{d\over dt}\langle\eta_{c}^{k}\rangle=-g\langle\sigma_{b}^{\dagger k}\hat{b}\rangle-g\langle\hat{b}^{\dagger}\hat\sigma_{b}^{k}\rangle,
\end{equation}
where
\begin{equation}\label{13}
\hat{\sigma}_{c}^{k}=|c\rangle_{k}~_{k}\langle a|,
\end{equation}
\begin{equation}\label{14}
\hat{\eta}_{a}^{k}=|a\rangle_{k}~_{k}\langle a|,
\end{equation}
\begin{equation}\label{15}
\hat{\eta}_{b}^{k}=|b\rangle_{k}~_{k}\langle b|,
\end{equation}
\begin{equation}\label{16}
\hat{\eta}_{c}^{k}=|c\rangle_{k}~_{k}\langle c|.
\end{equation}

We see that Eqs. (\ref{7})-(\ref{12}) are nonlinear differential equations and hence it is not possible to find exact time-dependent solutions of these equations. We intend to overcome this problem by applying the large-time approximation [12]. Then using this approximation scheme, we get from Eq. (\ref{5}) the approximately valid relation
\begin{equation}\label{17}
\hat{b}=-{2g\over\kappa}(\hat{\sigma}_{a}^{k}+\hat{\sigma}_{b}^{k}).
\end{equation}
Evidently, this turns out to be an exact relation at steady state.
Now introducing ~Eq. (\ref{17}) into Eqs. (\ref{7})-(\ref{12}), we get
\begin{equation}\label{18}
{d\over dt}\langle\hat{\sigma}_{a}^{k}\rangle=-\gamma_{c}\langle\hat{\sigma}_{a}^{k}\rangle,
\end{equation}
\begin{equation}\label{19}
{d\over dt}\langle\hat{\sigma}_{b}^{k}\rangle=-{\gamma_{c}\over 2}\langle\hat{\sigma}_{b}^{k}\rangle+\gamma_{c}\langle\hat{\sigma}_{a}^{k}\rangle,
\end{equation}
\begin{equation}\label{20}
{d\over dt}\langle\hat{\eta}_{a}^{k}\rangle=-\gamma_{c}\langle\hat{\eta}_{a}^{k}\rangle,
\end{equation}
\begin{equation}\label{22}
{d\over dt}\langle\hat{\eta}_{b}^{k}\rangle=-\gamma_{c}\langle\hat{\eta}_{b}^{k}\rangle+\gamma_{c}\langle\hat{\eta}_{a}^{k}\rangle,
\end{equation}
\begin{equation}\label{23}
{d\over dt}\langle\hat{\eta}_{c}^{k}\rangle=\gamma_{c}\langle\hat{\eta}_{b}^{k}\rangle,
\end{equation}
where
\begin{equation}\label{24}
\gamma_{c}={4g^{2}\over\kappa}.
\end{equation}
We prefer to call the parameter defined by Eq. (\ref{24}) the stimulated emission decay constant.

The three-level atoms available in the cavity are pumped from the bottom to the top level by means of electron bombardment. The pumping process must surely affect the dynamics of $\langle\hat{\eta}^{k}_{a}\rangle$ and $\langle\hat{\eta}^{k}_{c}\rangle$. If $r_{a}$ represents the rate at which a single atom is pumped from the bottom to the top level, then $\langle\hat{\eta}^{k}_{a}\rangle$ increases at the rate of $r_{a}\langle\hat{\eta}^{k}_{c}\rangle$ and $\langle\hat{\eta}^{k}_{c}\rangle$ decreases at the same rate. In view of this, we rewrite
Eqs. (\ref{20}) and (\ref{23}) as
\begin{equation}\label{25}
{d\over dt}\langle\hat{\eta}_{a}^{k}\rangle=-\gamma_{c}\langle\hat{\eta}_{a}^{k}\rangle+r_{a}\langle\hat{\eta}_{c}^{k}\rangle
\end{equation}
and
\begin{equation}\label{26}
{d\over dt}\langle\hat{\eta}_{c}^{k}\rangle=\gamma_{c}\langle\hat{\eta}_{b}^{k}\rangle-r_{a}\langle\hat{\eta}_{c}^{k}\rangle.
\end{equation}

We next sum Eqs. (\ref{18}), (\ref{19}), (\ref{22}), (\ref{25}), and (\ref{26}) over the $N$ three-level atoms, so that
\begin{equation}\label{36}
{d\over dt}\langle\hat{m}_{a}\rangle=-\gamma_{c}\langle\hat{m}_{a}\rangle,
\end{equation}
\begin{equation}\label{37}
{d\over dt}\langle\hat{m}_{b}\rangle=-{\gamma_{c}\over 2}\langle\hat{m}_{b}\rangle
+\gamma_{c}\langle\hat{m}_{a}\rangle,
\end{equation}
\begin{equation}\label{39}
{d\over dt}\langle\hat{N}_{a}\rangle=-\gamma_{c}\langle\hat{N}_{a}\rangle+r_{a}\langle\hat{N}_{c}\rangle,
\end{equation}
\begin{equation}\label{40}
{d\over dt}\langle\hat{N}_{b}\rangle=-\gamma_{c}\langle\hat{N}_{b}\rangle+\gamma_{c}\langle\hat{N}_{a}\rangle,
\end{equation}
\begin{equation}\label{41}
{d\over dt}\langle\hat{N}_{c}\rangle=\gamma_{c}\langle\hat{N}_{b}\rangle-r_{a}\langle\hat{N}_{c}\rangle,
\end{equation}
in which
\begin{equation}\label{42}
\hat{m}_{a}=\sum_{k=1}^{N}\hat{\sigma}_{a}^{k},
\end{equation}
\begin{equation}\label{43}
\hat{m}_{b}=\sum_{k=1}^{N}\hat{\sigma}_{b}^{k},
\end{equation}
\begin{equation}\label{45}
\hat{N}_{a}=\sum_{k=1}^{N}\hat{\eta}_{a}^{k},
\end{equation}
\begin{equation}\label{46}
\hat{N}_{b}=\sum_{k=1}^{N}\hat{\eta}_{b}^{k},
\end{equation}
\begin{equation}\label{47}
\hat{N}_{c}=\sum_{k=1}^{N}\hat{\eta}_{c}^{k},
\end{equation}
with the operators $\hat{N}_{a}$, $\hat{N}_{b}$, and $\hat{N}_{c}$ representing the number of atoms in the top, middle, and bottom levels.
In addition, employing the completeness relation
\begin{equation}\label{48}
\hat{\eta}_{a}^{k}+\hat{\eta}_{b}^{k}+\hat{\eta}_{c}^{k}=\hat{I},
\end{equation}
we easily arrive at
\begin{equation}\label{49}
\langle\hat{N}_{a}\rangle+\langle\hat{N}_{b}\rangle+\langle\hat{N}_{c}\rangle=N.
\end{equation}

Furthermore, using the definition given by (\ref{42}) and setting for any $k$
\begin{equation}\label{50}
\hat{\sigma}_{a}^{k}=|b\rangle\langle a|,
\end{equation}
we have
\begin{equation}\label{51}
\hat{m}_{a}=N|b\rangle\langle a|.
\end{equation}
Following the same procedure, one can also check that
\begin{equation}\label{52}
\hat{m}_{b}=N|c\rangle\langle b|,
\end{equation}
\begin{equation}\label{53}
\hat{m}_{c}=N|c\rangle\langle a|,
\end{equation}
\begin{equation}\label{54}
\hat{N}_{a}=N|a\rangle\langle a|,
\end{equation}
\begin{equation}\label{55}
\hat{N}_{b}=N|b\rangle\langle b|,
\end{equation}
\begin{equation}\label{56}
\hat{N}_{c}=N|c\rangle\langle c|,
\end{equation}
where
\begin{equation}
\hat{m}_{c}=\sum_{k=1}^{N}\hat{\sigma}_{c}^{k}.
\end{equation}
Moreover, using the definition
\begin{equation}\label{57}
\hat{m}=\hat{m}_{a}+\hat{m}_{b}
\end{equation}
and taking into account Eqs. (\ref{51})-(\ref{56}),
it can be readily established that
\begin{equation}\label{58}
\hat{m}^{\dagger}\hat{m}=N(\hat{N}_{a}+\hat{N}_{b}),
\end{equation}
\begin{equation}\label{59}
\hat{m}\hat{m}^{\dagger}=N(\hat{N}_{b}+\hat{N}_{c}),
\end{equation}
\begin{equation}\label{60}
\hat{m}^{2}=N\hat{m}_{c}.
\end{equation}

In the presence of $N$ three-level atoms, we rewrite Eq. (\ref{5}) as
\begin{equation}\label{61}
{d\hat{b}\over dt}=-{\kappa\over 2}\hat{b}+\lambda\hat{m},
\end{equation}
in which $\lambda$ is a constant whose value remains to be fixed. Applying the steady-state solution of
Eq. (\ref{5}), we get
\begin{equation}\label{62}
[\hat{b},\hat{b}^{\dagger}]_{k}={\gamma_{c}\over\kappa}(\hat{\eta}_{c}^{k}-\hat{\eta}_{a}^{k})
\end{equation}
and on summing over all atoms, we have
\begin{equation}\label{63}
[\hat{b},\hat{b}^{\dagger}]={\gamma_{c}\over\kappa}(\hat{N}_{c}-\hat{N}_{a}),
\end{equation}
where
\begin{equation}\label{64}
[\hat{b},\hat{b}^{\dagger}]=\sum_{k=1}^{N}[\hat{b},\hat{b}^{\dagger}]_{k}
\end{equation}
stands for the commutator of $\hat{b}$ and $\hat{b}^{\dagger}$ when the cavity mode is interacting with all the $N$ three-level atoms.
On the other hand, using the steady-state solution of Eq. (\ref{61}), one can easily verify that
\begin{equation}\label{65}
[\hat{b},\hat{b}^{\dagger}]=N\bigg({2\lambda\over\kappa}\bigg)^{2}(\hat{N}_{c}-\hat{N}_{a}).
\end{equation}
Thus on account of Eqs. (\ref{63}) and (\ref{65}), we see that
\begin{equation}\label{66}
\lambda={\pm}{g\over\sqrt{N}}
\end{equation}
and in view of this result, Eq. (\ref{61}) can be written as
\begin{equation}\label{67}
{d\hat{b}\over dt}=-{\kappa\over 2}\hat{b}+{g\over\sqrt{N}}\hat{m}.
\end{equation}

\section*{3. Photon statistics}
We next seek to calculate the mean and variance of the photon number at steady state. However, we need first to establish several important relations. Hence with the aid of (\ref{49}), one can put Eq. (\ref{39}) in the form
\begin{equation}\label{68}
{d\over dt}\langle\hat{N}_{a}\rangle=-(\gamma_{c}+r_{a})\langle\hat{N}_{a}\rangle+r_{a}(N-\langle\hat{N}_{b}\rangle)
\end{equation}
and application of the large-time approximation scheme to Eq. (\ref{40}) yields
\begin{equation}\label{69}
\langle\hat{N}_{b}\rangle=\langle\hat{N}_{a}\rangle.
\end{equation}
Thus on taking into account this result, we find the steady-state solution of Eq. (\ref{68}) to be
\begin{equation}\label{70}
\langle\hat{N}_{a}\rangle={r_{a}N\over{\gamma_{c}+2r_{a}}}.
\end{equation}
We also notice that at steady state Eq. (\ref{39}) leads to
\begin{equation}\label{71}
\langle\hat{N}_{c}\rangle={\gamma_{c}\over r_{a}}\langle\hat{N}_{a}\rangle.
\end{equation}

We now proceed to calculate the expectation value of the atomic operator $\hat{m}_{c}$. We assume that the state vector of a three-level atom, put in the form
\begin{equation}\label{71a}
|\psi\rangle_{k}=C_{a}|a\rangle_{k}+C_{b}|b\rangle_{k}+C_{c}|c\rangle_{k},
\end{equation}
can be used to determine the expectation value of an atomic operator formed by a pair of identical energy levels or by two distinct energy levels between which transition with the emission of a photon is dipole forbidden. One can thus readily establish that
\begin{equation}\label{71b}
C_{a}C_{a}^{*}=\langle\hat{\eta}_{a}^{\kappa}\rangle,
\end{equation}
\begin{equation}\label{71c}
C_{c}C_{c}^{*}=\langle\hat{\eta}_{c}^{\kappa}\rangle,
\end{equation}
and
\begin{equation}\label{71d}
\langle\hat{\sigma}_{c}^{k}\rangle=C_{a}C_{c}^{*}.
\end{equation}
In order to have a mathematically manageable analysis, we take $\langle\hat{\sigma}_{c}^{k}\rangle$ to be real, so that  \begin{equation}\label{71e}
C_{a}C_{c}^{*}=C_{a}^{*}C_{c}
\end{equation}
and on subtracting $C_{a}C_{c}$ from both sides of this equation, we have
\begin{equation}\label{71f}
C_{a}(C_{c}^{*}-C_{c})+C_{c}(C_{a}-C_{a}^{*})=0.
\end{equation}
Hence for all possible values of $C_{a}$ and $C_{c}$, we see that
\begin{equation}\label{71g}
C_{a}^{*}=C_{a}
\end{equation}
and
\begin{equation}\label{71h}
C_{c}^{*}=C_{c}.
\end{equation}
Now on account of (\ref{71b}) and (\ref{71c}) together with (\ref{71g}) and (\ref{71h}), Eq. (\ref{71d}) takes the form
\begin{equation}\label{71i}
\langle\hat{\sigma}_{c}^{k}\rangle=\sqrt{\langle\hat{\eta}_{a}^{k}\rangle\langle\hat{\eta}_{c}^{k}\rangle}.
\end{equation}
Finally on summing over $k$ from 1 up to $N$ and taking into account Eqs. (\ref{45}) and (\ref{47}), we have
\begin{equation}\label{71i}
\langle\hat{m}_{c}\rangle=\sqrt{\langle\hat{N}_{a}\rangle\langle\hat{N}_{c}\rangle}.
\end{equation}

Furthermore, adding Eqs. (\ref{36}) and (\ref{37}), we have
\begin{equation}\label{72}
{d\over dt}\langle\hat{m}\rangle=-{1\over 2}\gamma_{c}\langle\hat{m}\rangle+{1\over 2}\gamma_{c}\langle\hat{m}_{a}\rangle.
\end{equation}
In order to include the effect of the pumping process, we rewrite this equation as
\begin{equation}\label{73}
{d\hat{m}\over dt}=-{1\over 2}\mu\hat{m}
+{1\over 2}\mu\hat{m}_{a}+\hat{F}_{m}(t),
\end{equation}
in which $\hat{F}_{m}(t)$ is a noise operator with vanishing mean and $\mu$ is a parameter whose value remains to be determined. Employing the relation
\begin{equation}\label{74}
{d\over dt}\bigg\langle\hat{m}^{\dagger}\hat{m}\bigg\rangle=\bigg\langle{d\hat{m}^{\dagger}\over dt}\hat{m}\bigg\rangle
+\bigg\langle\hat{m}^{\dagger}{d\hat{m}\over dt}\bigg\rangle,
\end{equation}
along with Eq. (\ref{73}), we easily find
\begin{eqnarray}\label{75}
{d\over dt}\langle\hat{m}^{\dagger}\hat{m}\rangle=-\mu\langle\hat{m}^{\dagger}\hat{m}\rangle
+\mu\langle\hat{m}^{\dagger}_{a}\hat{m}_{a}\rangle
+\langle\hat{F}^{\dagger}_{m}(t)\hat{m}\rangle
+\langle\hat{m}^{\dagger}\hat{F}_{m}(t)\rangle,
\end{eqnarray}
from which follows
\begin{eqnarray}\label{76}
{d\over dt}\langle\hat{N}_{a}+\hat{N}_{b}\rangle=-\mu\langle\hat{N}_{a}+\hat{N}_{b}\rangle
+\mu\langle\hat{N}_{a}\rangle
+{1\over N}\langle\hat{F}^{\dagger}_{m}(t)\hat{m}\rangle
+{1\over N}\langle\hat{m}^{\dagger}\hat{F}_{m}(t)\rangle.
\end{eqnarray}
On the other hand, on account of Eqs. (\ref{39}) and (\ref{40}), we see that
\begin{equation}\label{77}
{d\over dt}\langle\hat{N}_{a}+\hat{N}_{b}\rangle=-\gamma_{c}\langle\hat{N}_{b}\rangle
+r_{a}\langle\hat{N}_{c}\rangle.
\end{equation}
This can be rewritten as
\begin{eqnarray}\label{78}
{d\over dt}\langle\hat{N}_{a}+\hat{N}_{b}\rangle=-(\gamma_{c}+2r_{a})\langle\hat{N}_{a}
+\hat{N}_{b}\rangle
+(\gamma_{c}+2r_{a})\langle\hat{N}_{a}\rangle+r_{a}N.
\end{eqnarray}
Hence comparison of Eqs. (\ref{76}) and (\ref{78}) shows that
\begin{equation}\label{79}
\mu=\gamma_{c}+2r_{a}
\end{equation}
and
\begin{equation}\label{80}
\langle\hat{F}^{\dagger}_{m}(t)\hat{m}\rangle+\langle\hat{m}^{\dagger}\hat{F}_{m}(t)\rangle=r_{a}N^{2}.
\end{equation}
We observe that Eq. (\ref{80}) is equivalent to
\begin{equation}\label{81}
\langle\hat{F}_{m}^{\dagger}(t)\hat{F}_{m}(t')\rangle=r_{a}N^{2}\delta(t-t').
\end{equation}
Upon casting Eq. (\ref{36}) into the form
\begin{equation}\label{82}
{d\over dt}\hat{m}_{a}=-{1\over 2}\mu\hat{m}_{a}+\hat{F}_{a}(t),
\end{equation}
one can also easily verify that $\mu$ has the value given by (\ref{79}) and
\begin{equation}\label{83}
\langle\hat{F}_{a}^{\dagger}(t)\hat{F}_{a}(t')\rangle=r_{a}N^{2}\delta(t-t').
\end{equation}

With the atoms considered to be initially in the bottom level, the expectation value of the solution of Eq. (\ref{82}) happens to be
\begin{equation}\label{84}
\langle\hat{m}_{a}(t)\rangle=0.
\end{equation}
Then on account of this result, the expectation value of the solution of Eq. (\ref{73}) turns out to be
\begin{equation}\label{85}
\langle\hat{m}(t)\rangle=0.
\end{equation}
On the other hand, the expectation value of
the solution of Eq. (\ref{67}) is expressible as
\begin{eqnarray}\label{86}
\langle\hat{b}(t)\rangle=\langle\hat{b}(0)\rangle e^{-\kappa t/2}
+{g\over\sqrt{N}}e^{-\kappa t/2}
\int_{0}^{t}e^{\kappa t'/2}\langle\hat{m}(t')\rangle dt'.
\end{eqnarray}
Now in view of Eq. (\ref{85}) and the assumption that the cavity light is initially in a vacuum state, Eq. (\ref{86}) goes over into
\begin{equation}\label{87}
\langle\hat{b}(t)\rangle=0.
\end{equation}
We observe on the basis of Eqs. (\ref{67}) and (\ref{87}) that $\hat{b}$ is a Gaussian variable with zero mean.
The steady-state solution of Eq. (\ref{67}) is given by
\begin{equation}\label{88}
\hat{b}={2g\over\kappa\sqrt{N}}\hat{m}.
\end{equation}
Now using Eq. (\ref{88}) along with (\ref{58}), the mean photon number of the cavity light is expressible as
\begin{equation}\label{89}
\langle\hat{b}^{\dagger}\hat{b}\rangle={\gamma_{c}\over\kappa}\bigg(\langle\hat{N}_{a}\rangle+\langle\hat{N}_{b}\rangle\bigg).
\end{equation}

It proves to be convenient to refer to the regime of laser operation with more atoms in the top level than in the bottom level as above threshold, the regime of laser operation with equal number of atoms in the top and bottom levels as threshold, and the regime of laser operation with less atoms in the top level than in the bottom level as below threshold. Thus according to Eq. (\ref{71}) for the laser operating above threshold $\gamma_{c}<r_{a}$, for the laser operating at threshold $\gamma_{c}=r_{a}$, and for the laser operating below threshold $\gamma_{c}>r_{a}$.
We note that for the laser operating well above threshold, Eq. (\ref{89}) reduces to
\begin{equation}\label{90}
\bar{n}={\gamma_{c}\over\kappa}N
\end{equation}
and for  the laser operating at threshold, we find
\begin{equation}\label{91}
\bar{n}={2\gamma_{c}\over 3\kappa}N,
\end{equation}
in which $\bar{n}=\langle\hat{b}^{\dagger}\hat{b}\rangle$.

The variance of the photon number for the cavity light is expressible as
\begin{equation}\label{92}
(\Delta n)^{2}=\langle\hat{b}^{\dagger}\hat{b}\hat{b}^{\dagger}\hat{b}\rangle
-\langle\hat{b}^{\dagger}\hat{b}\rangle^{2} \end{equation}
and using the fact that $\hat{b}$ is a Gaussian variable with zero mean , we readily get
\begin{equation}\label{93}
(\Delta n)^{2}=\langle\hat{b}^{\dagger}\hat{b}\rangle\langle\hat{b}\hat{b}^{\dagger}\rangle+\langle\hat{b}^{\dagger 2}\rangle\langle\hat{b}^{2}\rangle.
\end{equation}
Employing once more Eq. (\ref{88}) and taking into account (\ref{59}), we find
\begin{equation}\label{94}
\langle\hat{b}\hat{b}^{\dagger}\rangle={\gamma_{c}\over\kappa}\bigg(\langle\hat{N}_{b}\rangle
+\langle\hat{N}_{c}\rangle\bigg).
\end{equation}
In addition, taking into account (\ref{71}), one can express Eq. (\ref{71i}) in the form
\begin{equation}\label{95}
\langle\hat{m}_{c}\rangle={\sqrt{\gamma_{c}\over r_{a}}}\langle\hat{N}_{a}\rangle.
\end{equation}
Hence with the aid of Eqs. (\ref{88}), (\ref{60}), and (\ref{95}), we easily get
\begin{equation}\label{96}
\langle\hat{b}^{2}\rangle={\gamma_{c}\over\kappa}\sqrt{\gamma_{c}\over r_{a}}\langle\hat{N}_{a}\rangle.
\end{equation}
Now on account of Eqs. (\ref{93}), (\ref{94}), and (\ref{96}) along with (\ref{69}) and (\ref{71}), we arrive at
\begin{equation}\label{98}
(\Delta n)^{2}={1\over 4}\overline{n}^{2}(2\eta+\eta+2),
\end{equation}
with $\eta=\gamma_{c}/r_{a}$.
We then see that for the laser operating well above threshold
\begin{equation}\label{99}
(\Delta n)^{2}={1\over 2}\overline{n}^{2},
\end{equation}
where $\overline{n}$ is given by Eq. (\ref{90}).
And for the laser operating at threshold, we get
\begin{equation}\label{100}
(\Delta n)^{2}={5\over 4}\overline{n}^{2},
\end{equation}
in which $\overline{n}$ given by Eq. (\ref{91}).

\section*{4. Quadrature squeezing}
In this section we wish to calculate the quadrature squeezing of the cavity and output light in any frequency interval.
\subsection*{4.1 Global quadrature squeezing}
The squeezing properties of the cavity light are described by two quadrature operators defined by
\begin{equation}\label{101}
\hat{b}_{+}=\hat{b}^{\dagger}+\hat{b}
\end{equation}
and
\begin{equation}\label{102}
\hat{b}_{-}=i(\hat{b}^{\dagger}-\hat{b}).
\end{equation}
It can be readily established that
\begin{equation}\label{103}
[\hat{b}_{-},\hat{b}_{+}]=2i{\gamma_{c}\over\kappa}(\hat{N}_{a}-\hat{N}_{c}).
\end{equation}
It then follows that
\begin{equation}\label{104}
\Delta b_{+}\Delta b_{-}\geq{\gamma_{c}\over\kappa}\bigg|\langle\hat{N}_{a}\rangle-\langle\hat{N}_{c}\rangle\bigg|.
\end{equation}
The variance of the quadrature operators is expressible as
\begin{equation}\label{105}
(\Delta b_{\pm})^{2}={\pm}\langle[\hat{b}^{\dagger}{\pm}\hat{b}]^{2}\rangle
{\mp}[\langle\hat{b}^{\dagger}\rangle{\pm}\langle\hat{b}\rangle]^{2},
\end{equation}
so that on account of (\ref{87}), we have
\begin{equation}\label{106}
(\Delta b_{\pm})^{2}=\langle\hat{b}^{\dagger}\hat{b}\rangle+\langle\hat{b}\hat{b}^{\dagger}\rangle
{\pm}\langle\hat{b}^{\dagger 2}\rangle{\pm}\langle\hat{b}^{2}\rangle.
\end{equation}
Now employing (\ref{89}), (\ref{94}), and (\ref{96}), we arrive at
\begin{equation}\label{107}
\hspace*{-7mm}(\Delta b_{+})^{2}={\gamma_{c}\over\kappa}\bigg(N+\langle\hat{N}_{a}\rangle
+2\sqrt{\gamma_{c}\over r_{a}}\langle\hat{N}_{a}\rangle\bigg)
\end{equation}
and
\begin{equation}\label{108}
\hspace*{-7mm}(\Delta b_{-})^{2}={\gamma_{c}\over\kappa}\bigg(N+\langle\hat{N}_{a}\rangle
-2\sqrt{\gamma_{c}\over r_{a}}\langle\hat{N}_{a}\rangle\bigg).
\end{equation}

We recall that the light generated by a two-level laser operating well above threshold is
coherent, the quadrature variance of which is given by [12]
\begin{equation}\label{109}
(\Delta b_{\pm})^{2}_{c}= {\gamma_{c}\over\kappa}N.
\end{equation}
We calculate the quadrature squeezing of the cavity (output) light relative to the quadrature variance of the cavity (output) coherent light. We then define the quadrature squeezing of the cavity light by
\begin{equation}\label{110}
S={(\Delta b_{\pm})^{2}_{c}-(\Delta b_{-})^{2}\over{(\Delta b_{\pm})^{2}_{c}}}.
\end{equation}
Hence employing (\ref{108}), (\ref{109}), and (\ref{70}), one can put Eq. (\ref{110}) in the form
\begin{equation}\label{111}
S={{2\sqrt{\eta}-1}\over{\eta+2}}.
\end{equation}
\begin{figure}[hbt]
\centering
\includegraphics[height=8cm,keepaspectratio]{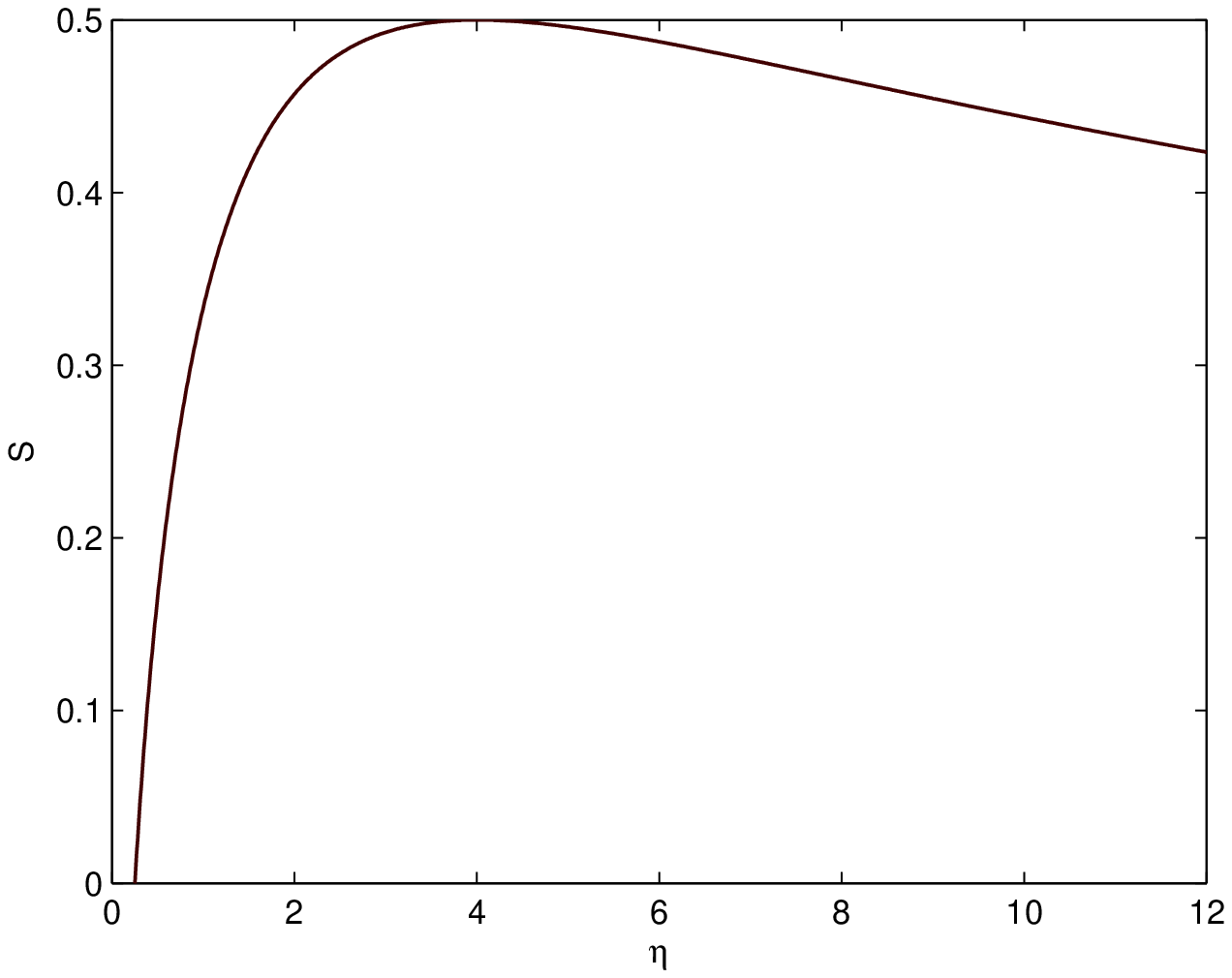}
\begin{center}
{\footnotesize{\bf Fig. 1}~~~~~A plot of Eq. (\ref{111}) versus $\eta$.}
\end{center}
\end{figure}
\noindent
We observe that, unlike the mean photon number, the quadrature squeezing does not depend on the number of atoms. This implies that the quadrature squeezing of the cavity light is independent of the number of photons.
The plot in Fig. 1 shows that the maximum squeezing of the cavity light is 50 $\%$ below the coherent-state level and this occurs when the three-level laser is operating below threshold at $\gamma_{c}=4r_{a}$.

On the other hand, we define the quadrature squeezing of the output light by
\begin{equation}\label{112}
S^{out}={(\Delta b^{out}_{\pm})^{2}_{c}-(\Delta b_{-}^{out})^{2}\over(\Delta b^{out}_{\pm})^{2}_{c}},
\end{equation}
where $(\Delta b^{out}_{\pm})^{2}_{c}$ is the quadrature variance of the output coherent light.
Since all calculations in this analysis are carried out by putting the vacuum noise operators in normal order,
one can write
\begin{equation}\label{113}
\hat{b}^{out}=\sqrt{\kappa}\hat{b}.
\end{equation}
We then easily see that
\begin{equation}\label{114}
(\Delta b^{out}_{\pm})^{2}_{c}=\kappa(\Delta b_{\pm})^{2}_{c}
\end{equation}
and
\begin{equation}\label{115}
(\Delta b^{out}_{-})^{2}=\kappa(\Delta b_{-})^{2}.
\end{equation}
Now in view of Eqs. (\ref{112}), (\ref{114}), (\ref{115}), and (\ref{110}), we arrive at
\begin{equation}\label{116}
S^{out}=S.
\end{equation}
We observe that the quadrature squeezing of the output light is equal to that of the cavity light.

\subsection*{4.2 Local quadrature squeezing}
We finally seek to obtain the quadrature squeezing of the cavity (output) light in a given frequency interval. To this end, we first determine the spectrum of quadrature fluctuations of the cavity light. We define this spectrum for a single-mode light with central frequency $\omega_{0}$ by
\begin{eqnarray}\label{117} S_{\pm}(\omega)={1\over\pi}Re\int_{0}^{\infty}\left\langle\hat{b}_{\pm}(t),
\hat{b}_{\pm}(t+\tau)\right\rangle_{ss}
e^{i(\omega-\omega_{0})\tau}d\tau.
\end{eqnarray}
Upon integrating both sides of Eq. (\ref{117}) over $\omega$, we get
\begin{equation}\label{118}
\int_{-\infty}^{\infty} S_{\pm}(\omega)d\omega=(\Delta b_{\pm})^{2},
\end{equation}
in which
\begin{equation}\label{119}
(\Delta b_{\pm})^{2}=\langle\hat{b}_{\pm}(t), \hat{b}_{\pm}(t)\rangle_{ss}
\end{equation}
is the quadrature variance of the light mode at steady state.
On the basis of the result given by Eq. (\ref{118}), we assert that $S_{\pm}(\omega)d\omega$ is the steady-state quadrature variance of the light mode in the interval between $\omega$ and $\omega+d\omega$.
In view of Eq. (\ref{87}), we note that
\begin{equation}\label{120}
\langle\hat{b}_{\pm}(t),\hat{b}_{\pm}(t+\tau)\rangle=\langle\hat{b}_{\pm}(t)
\hat{b}_{\pm}(t+\tau)\rangle.
\end{equation}

We now proceed to determine the two-time correlation function that appears in Eq. (\ref{120}) for the cavity light.
To this end, we realize that the solution of Eq. (\ref{67}) can also be written as
\begin{eqnarray}\label{121}
\hat{b}(t+\tau)=\hat{b}(t)e^{-\kappa \tau/2}
+{g\over\sqrt{N}}e^{-\kappa \tau/2}
\int_{0}^{\tau}e^{\kappa\tau'/2}\hat{m}(t+\tau')d\tau'.
\end{eqnarray}
On the other hand, the solution of Eq. (\ref{73}) is expressible as
\begin{eqnarray}\label{122}
\hat{m}(t+\tau)=\hat{m}(t)e^{-\mu\tau/2}
+ e^{-\mu\tau/2}\int_{0}^{\tau}e^{\mu\tau''/2}
\bigg({1\over 2}\mu\hat{m}_{a}(t+\tau'')
+\hat{F}_{m}(t+\tau'')\bigg)d\tau''.
\end{eqnarray}
Applying the large-time approximation scheme to Eq. (\ref{82}), we get
\begin{equation}\label{123}
\hat{m}_{a}(t+\tau)={2\over\mu}\hat{F}_{a}(t+\tau),
\end{equation}
so that on introducing this into Eq. (\ref{122}), there follows
\begin{eqnarray}\label{124}
\hat{m}(t+\tau)=\hat{m}(t)e^{-\mu\tau/2}
+e^{-\mu\tau/2}\int_{0}^{\tau}e^{\mu\tau''/2}
\bigg(\hat{F}_{a}(t+\tau'')
+\hat{F}_{m}(t+\tau'')\bigg)d\tau''.
\end{eqnarray}
Now combination of Eqs. (\ref{121}) and (\ref{124}) yields
\begin{eqnarray}\label{125}
\hspace*{-5mm}\hat{b}(t+\tau)\hspace*{-3mm}&=\hspace*{-3mm}&\hat{b}(t)e^{-\kappa \tau/2}
+{2g\hat{m}(t)\over{\sqrt{N}(\kappa-\mu)}}
\bigg[e^{-\mu\tau/2}-e^{-\kappa\tau/2}\bigg]
+{g\over{\sqrt{N}}}e^{-\kappa\tau/2}\int_{0}^{\tau}d\tau'e^{(\kappa-\mu)\tau'/2}\nonumber\\&&
\times\int_{0}^{\tau'}d\tau''e^{\mu\tau''/2}\bigg(\hat{F}_{a}(t+\tau'')
+\hat{F}_{m}(t+\tau'')\bigg).
\end{eqnarray}
Hence using this result, we readily arrive at
\begin{eqnarray}\label{126}
\langle\hat{b}^{\dagger}(t)\hat{b}(t+\tau)\rangle=\langle\hat{b}^{\dagger}(t)\hat{b}(t)\rangle
e^{-\kappa\tau/2}
+{2g\over{\sqrt{N}(\kappa-\mu)}}\langle\hat{b}^{\dagger}(t)\hat{m}(t)\rangle
[e^{-\mu\tau/2}-e^{-\kappa\tau/2}].
\end{eqnarray}
Applying once more the large-time approximation, one gets from Eq. (\ref{67})
\begin{equation}\label{127}
\hat{m}(t)={\kappa\sqrt{N}\over 2g}\hat{b}(t).
\end{equation}
With this substituted into Eq. (\ref{126}), there emerges
\begin{eqnarray}\label{128}
\langle\hat{b}^{\dagger}(t)\hat{b}(t+\tau)\rangle=\langle\hat{b}^{\dagger}(t)\hat{b}(t)\rangle
\bigg({\kappa\over{\kappa-\mu}}e^{-\mu\tau/2}
-{\mu\over{\kappa-\mu}}e^{-\kappa\tau/2}\bigg).
\end{eqnarray}
Following a similar procedure, one can also readily  establish that
\begin{eqnarray}\label{129}
\langle\hat{b}(t)\hat{b}^{\dagger}(t+\tau)\rangle=\langle\hat{b}(t)\hat{b}^{\dagger}(t)\rangle
\bigg({\kappa\over{\kappa-\mu}}e^{-\mu\tau/2}
-{\mu\over{\kappa-\mu}}e^{-\kappa\tau/2}\bigg)\end{eqnarray}
and
\begin{eqnarray}\label{130}
\langle\hat{b}(t)\hat{b}(t+\tau)\rangle=\langle\hat{b}^{2}(t)
\rangle\bigg({\kappa\over{\kappa-\mu}}e^{-\mu\tau/2}
-{\mu\over{\kappa-\mu}}e^{-\kappa\tau/2}\bigg).
\end{eqnarray}
Therefore, on account of  Eqs. (\ref{128}), (\ref{129}), and (\ref{130}), there follows
\begin{eqnarray}\label{131}
\langle\hat{b}_{\pm}(t),\hat{b}_{\pm}(t+\tau)\rangle_{ss}=(\Delta b_{\pm})^{2}\bigg({\kappa\over\kappa-\mu}
e^{-\mu\tau/2}-{\mu\over\kappa-\mu}e^{-\kappa\tau/2}\bigg).
\end{eqnarray}
Now on introducing Eq. (\ref{131}) into Eq. (\ref{117}) and carrying out the integration, we find the spectrum of the minus quadrature fluctuations for the cavity light to be
\begin{eqnarray}\label{132}
S_{-}(\omega)=(\Delta b_{-})^{2}\bigg[{\kappa\over{\kappa-\mu}}\bigg({\mu/2\pi\over{(\omega-\omega_{0})^{2}
+[\mu/2]^{2}}}\bigg)
-{\mu\over{\kappa-\mu}}\bigg({\kappa/2\pi\over{(\omega-\omega_{0})^{2}+[\kappa/2]^{2}}}\bigg)\bigg],
\end{eqnarray}
in which $(\Delta b_{-})^{2}$ is given by (\ref{108}).

We realize that the variance of the minus quadrature in the interval between $\omega'=-\lambda$ and $\omega'=\lambda$ is expressible as
\begin{equation}\label{133}
(\Delta b_{-})^{2}_{{\pm}\lambda}=\int_{-\lambda}^{\lambda}S_{-}(\omega')d\omega',
\end{equation}
in which $\omega'=\omega-\omega_{0}.$
Hence taking into account Eq. (\ref{132}), we readily get
\begin{equation}\label{134}
(\Delta b_{-})_{{\pm}\lambda}^{2}=z(\lambda)(\Delta b_{-})^{2},
\end{equation}
where $z(\lambda)$ is given by
\begin{eqnarray}\label{135}
z(\lambda)={2\kappa/\pi\over{\kappa-\mu}}tan^{-1}\bigg({2\lambda\over\mu}\bigg)
-{2\mu/\pi\over{\kappa-\mu}}tan^{-1}\bigg({2\lambda\over\kappa}\bigg).
\end{eqnarray}
On account of the result described by (\ref{134}), the quadrature variance of the output coherent light can be written in the same frequency interval as
\begin{equation}\label{136}
(\Delta b_{\pm})_{c{\pm}\lambda}^{2}=z(\lambda)(\Delta b_{\pm})_{c}^{2}.
\end{equation}
We define the quadrature squeezing of the cavity light in the $\lambda_{\pm}$ frequency interval by
\begin{equation}\label{137}
S_{{\pm}\lambda}={(\Delta b_{\pm})^{2}_{{c}{\pm}\lambda}-(\Delta b_{-})^2_{{\pm}\lambda}\over{(\Delta b_{\pm})^{2}_{{c}{\pm}\lambda}}},
\end{equation}
so that on account of (\ref{134}), (\ref{136}), and  (\ref{110}),
there follows
\begin{equation}
S_{{\pm}\lambda}=S.
\end{equation}
This shows that the quadrature squeezing of the cavity light in a given frequency interval is equal to that of the cavity light in the entire frequency interval.

Finally, defining the quadrature squeezing of the output light in the aforementioned frequency interval by
\begin{equation}\label{138}
S^{out}_{{\pm}\lambda}={(\Delta b^{out}_{\pm})_{c{\pm}\lambda}^{2}-(\Delta b^{out}_{-})^2_{{\pm}\lambda}\over{(\Delta b^{out}_{\pm})_{c{\pm}\lambda}^{2}}}
\end{equation}
and taking into account the fact that
\begin{equation}\label{139}
(\Delta b^{out}_{\pm})^{2}_{{c}{{\pm}\lambda}}=z(\lambda)(\Delta b^{out}_{\pm})_{c}
\end{equation}
and
\begin{equation}\label{140}
(\Delta b^{out}_{-})_{{\pm}\lambda}^{2}=z(\lambda)(\Delta b^{out}_{-}),
\end{equation}
we arrive at
\begin{equation}\label{141}
S^{out}_{{\pm}\lambda}=S^{out}.
\end{equation}
We see that the quadrature squeezing of the output light in a certain frequency interval is the same as that of the output light in the entire frequency interval.

\section*{7. Conclusion}
We have considered a three-level laser in which the three-level atoms available in a closed cavity are pumped from the bottom to the top level by means of electron bombardment.
We have carried out our analysis by putting the vacuum noise operators in normal order and applying the large-time approximation scheme. The procedure of normal ordering the noise operators renders the vacuum reservoir to be a noiseless physical entity. We maintain the standpoint that the notion of a noiseless vacuum reservoir would turn out to be compatible with observation. Based on the definition given by Eq. (\ref{24}), we infer that an atom in the top or middle level and inside a closed cavity emits a photon due to its interaction with the cavity light. We certainly identify this process to be stimulated emission.

We have seen that the light generated by the three-level laser operating under the condition $\gamma_{c}>{1\over 4}r_{a}$ is in a squeezed state, with the maximum quadrature squeezing being $50\%$ below the coherent-state level. This occurs when the three-level laser is operating below threshold at $\gamma_{c}=4r_{a}$.
We have also established that the quadrature squeezing of the laser light has the same value in any frequency interval. On the basis of this result, we come to the conclusion that the quadrature squeezing of the laser light is an intrinsic property of the individual photons.

\vspace*{5mm}
\noindent
{\bf References}
\vspace*{3mm}

\noindent
[1] M.O. Scully, K. Wodkiewicz, M.S. Zubairy, J. Bergou, N. Lu, J. Meyer ter Vehn, Phys. Rev. Lett. \hspace*{6mm}60, 1832 (1988).\newline
[2] Fesseha Kassahun, Fundamentals of Quantum  Optics (Lulu Press Inc., North Carolina, 2010).\newline
[3] N. Lu, F.X. Zhao, J. Bergou, Phys. Rev. A 39, 5189 (1989).\newline
[4] Eyob Alebachew and K. Fesseha, Opt. Commun. 265, 314 (2006).\newline
[5] M.O. Scully and M.S. Zubairy, Opt. Commun. 66, 303 (1988).\newline
[6] N. Lu and S.Y. Zhu, Phys. Rev. A 41, 2865 (1990).\newline
[7] N.A. Ansari, J. Gea-Banacloche, and M.S. Zubairy,  Phys. Rev. A 41, 5179 (1990).\newline
[8] C.A. Blockely and D.F. Walls, Phys. Rev. A 43, 5049 (1991).\newline
[9] N.A. Ansari, Phys. Rev. A 48, 4686 (1993).\newline
[10] J. Anwar and M.S. Zubairy, Phys. Rev. A 49, 481 (1994).\newline
[11] K. Fesseha, Phys. Rev. A 63, 33811 (2001).\newline
[12] Fesseha Kassahun, Opt. Commun. 284, 1357 (2011).\newline

\end{document}